\documentclass[%
 reprint,
superscriptaddress,
 amsmath,amssymb,
 aps,
prc,
floatfix,
]{revtex4-1}
\usepackage{subfigure}
\usepackage{color,graphicx}
\usepackage{amsmath,amssymb,longtable}
\usepackage[%
bookmarks=true,%
bookmarksnumbered=true,%
bookmarkstype=toc,%
colorlinks=true,%
linkcolor=blue,%
citecolor=blue,%
]{hyperref}
\usepackage{feynmp-auto}

\newcommand{\bra}{\langle}
\newcommand{\ket}{\rangle}

\begin{document}


\title{Ground-state properties of doubly magic nuclei from the unitary-model-operator approach
 with the chiral
 two- and three-nucleon forces}


\author{T. Miyagi}
\affiliation{Center for Nuclear Study, The University of Tokyo, 7-3-1 Hongo, Bunkyo, Tokyo 113-0033, Japan}
\affiliation{TRIUMF, 4004 Wesbrook Mall, Vancouver, British Columbia, V6T 2A3 Canada}
\author{T. Abe}
\affiliation{Department of Physics, The University of Tokyo, 7-3-1, Hongo, Bunkyo, Tokyo 113-0033, Japan}
\affiliation{Center for Nuclear Study, The University of Tokyo, 7-3-1 Hongo, Bunkyo, Tokyo 113-0033, Japan}
\author{M. Kohno}
\affiliation{Research Center for Nuclear Physics, Osaka University, Japan}
\author{P. Navr\'atil}
\affiliation{TRIUMF, 4004 Wesbrook Mall, Vancouver, British Columbia, V6T 2A3 Canada}
\author{R. Okamoto}
\affiliation{Senior Academy, Kyushu Institute of Technology, Tobata, Kitakyushu 804-0015, Japan}
\author{T. Otsuka}
\affiliation{RIKEN Nishina Center, 2-1 Hirosawa, Wako, Saitama 351-0198, Japan}
\affiliation{Department of Physics, The University of Tokyo, 7-3-1, Hongo, Bunkyo, Tokyo 113-0033, Japan}
\affiliation{National Superconducting Cyclotron Laboratory, Michigan State University, East Lansing, Michigan 48824, USA}
\affiliation{Instituut voor Kern- en Stralingsfysica, KU Leuven, B-3001 Leuven, Belgium}
\author{N. Shimizu}
\affiliation{Center for Nuclear Study, The University of Tokyo, 7-3-1 Hongo, Bunkyo, Tokyo 113-0033, Japan}
\author{S. R. Stroberg}
\affiliation{TRIUMF, 4004 Wesbrook Mall, Vancouver, British Columbia, V6T 2A3 Canada}
\affiliation{Physics Department, Reed College, Portland, Oregon, 97202 USA}

\date{\today}

\begin{abstract}
  The ground-state energies and radii for $^{4}$He, $^{16}$O, and
   $^{40}$Ca are calculated with the unitary-model-operator approach (UMOA).
  In the present study, we employ the similarity renormalization group (SRG) evolved
   nucleon-nucleon ($NN$) and three-nucleon ($3N$) interactions based on the chiral effective field theory.
  This is the first UMOA calculation with both $NN$ and $3N$ interactions.
  The calculated ground-state energies and radii are consistent with the recent {\it ab initio}
   results with the same interaction.
  We evaluate the expectation values with two- and three-body SRG
   evolved radius operators, in addition to those with the bare radius operator.
  With the aid of the higher-body evolution of radius operator,
   it is seen that the calculated radii tend to be SRG resolution-scale independent.
  We find that the SRG evolution gives minor modifications for the radius operator.
\end{abstract}


\maketitle

\section{Introduction}
Recent nuclear {\em ab initio} studies are encouraged by the development, in particular, of the
 nuclear interactions from the chiral effective field theory ($\chi$EFT)~\cite{Epelbaum2009, Machleidt2011}.
In the $\chi$EFT, nuclear interactions are obtained through the perturbation expansion
 of the chiral Lagrangian which
 is the effective Lagrangian of the quantum chromodynamics.
By taking into account the higher-order expansion terms, the systematic improvement of the nuclear interactions
 can be expected, for recent example, see Refs.~\cite{Binder2016, Entem2017, Binder2018}.
As another advantage of employing the $\chi$EFT, the consideration of higer-order terms in the perturbation series
 allows us the systematic derivation of the three-nucleon ($3N$) interaction.
With the development of the $\chi$EFT interactions, the impacts of the $3N$ force on nuclear structure
 calculation have been discussed extensively, for example,
 in light nuclei~\cite{Navratil2007, Hupin2014, Langhammer2015, Hupin2015},
 medium-mass nuclei~\cite{Otsuka2010, Hagen2012a, Hagen2012, Hergert2013a, Hergert2013,
 Hergert2014, Soma2014, Cipollone2015, Binder2014a},
 and infinite nuclear matter~\cite{Hebeler2011, Kohno2013, Hagen2014a, Tews2016}.

Besides the progress in nuclear forces, the advancement in many-body method is also necessary.
To deal with nuclear many-body problems, one can use the {\em ab initio} calculation methods
 such as no-core shell model (NCSM)~\cite{Barrett2013},
 quantum Monte Carlo methods~\cite{Carlson2015},
 nuclear lattice EFT calculations~\cite{Lee2009},
 coupled-cluster method~\cite{Hagen2014},
 self-consistent Green's function method~\cite{Dickhoff2004},
 in-medium similarity renormalization group approach~\cite{Hergert2016},
 and many-body perturbation theory~\cite{Tichai2016, Hu2016}.
Over the past decade, the tremendous advancements were made in nuclear {\em ab initio} studies.
Nowadays, the capability of the {\em ab initio} calculations has reached
 mass region $A \sim 100$~\cite{Binder2014a, Tichai2016, Morris2018}.
In addition to these methods, we can apply the unitary-model-operator approach
 (UMOA)~\cite{Suzuki1994, Miyagi2017a} to solve the nuclear many-body Schr\"odinger equation.
In the UMOA, a unitary transformation of the Hamiltonian is constructed so that the
 the one-particle-one-hole and two-particle-two-hole excitations are limitated from the
 transformed Hamiltonian.
So far, we calculated the ground-state energies and radii for some closed shell nuclei with only the
 nucleon-nucleon ($NN$) interactions.
In this work, we include the $3N$ interaction effect to the UMOA calculation for the first time.

Due to the non-perturbative nature of the nuclear force, in most cases,
 it is not possible to apply directly
 the nuclear interactions to the many-body calculations.
To bridge the gap between nuclear forces and many-body calculations,
 we evolve the nuclear Hamiltonian with the similarity
 renormalization group (SRG) flow equation~\cite{Bogner2007}.
Through the SRG evolution, we obtain the
 Hamiltonian whose coupling between low- and high-momentum regions is suppressed.
With such nuclear interactions,
 recent {\em ab initio} results significantly underestimate the nuclear radii, see for instance
 Refs.~\cite{Lapoux2016, Cipollone2015, Soma2014, Hergert2016}.
Since the nuclear size can affect the single-particle level structure of a nucleus,
 the reproduction of nuclear radii is one of the fundamental issues to discuss the nuclear structure.
As seen in NCSM calculations for few-body systems~\cite{Schuster2014, Schuster2015},
 we should also evolve consistently other operators than the Hamiltonian.
In this work, we demonstrate the effect of the SRG evolution on radius operator.

This paper is organized as follows.
The Hamiltonian and radius operators employed in this work are introduced in Sec.~\ref{sec:hr}.
Section~\ref{sec:umoa} describes briefly the formalism of the UMOA.
In Sec.~\ref{sec:res}, the numerical results for ${}^{4}$He, ${}^{16}$O, and ${}^{40}$Ca are given.
After comfirming convergence and consistency with the other {\em ab initio} results,
 the effects of the SRG evolution on radius operator are discussed.
The summary of the present work is given in Sec.~\ref{sec:conc}.

\section{Hamiltonian and Radius Operators \label{sec:hr}}
Our starting Hamiltonian is composed of the kinetic, $NN$, and $3N$ terms:
\begin{equation}
  \label{eq:hini}
  H = T + V^{NN} + V^{3N}.
\end{equation}
Here, $T$ is the kinetic energy operator.
The $V^{NN}$ and $V^{3N}$ indicate the $NN$ and $3N$ interactions, respectively.
Usually, the bare Hamiltonian $H$ is too
 "hard" to apply for many-body calculation methods.
It causes
 the slow convergence with respect to the size of model space
 and calculations demand the huge amount of computational resources.
To obtain the converged results from the feasible model-space calculations,
 the similarity-renormalization group (SRG) evolution~\cite{Bogner2007}
 is employed in this work.
We consider the unitary transformation of the original Hamiltonian:
\begin{equation}
  \label{eq:h_srg}
  H(\alpha) = U_{\rm SRG}^{\dag}(\alpha) H U_{\rm SRG}(\alpha).
\end{equation}
Here, $U_{\rm SRG}(\alpha)$ is the unitary transformation operator and
 is evolved by the flow equation:
\begin{equation}
  \label{eq:uevo}
  \frac{d U_{\rm SRG}(\alpha)}{d \alpha} = U_{\rm SRG}(\alpha) \eta(\alpha).
\end{equation}
The $\alpha$ is the resolution scale parameter of the flow equation in unit of
 fm$^{4}$.
The $\eta$ is called the generator of the SRG evolution
 and is taken as $\eta(\alpha) = [T, H(\alpha)]$.
Note that the initial condition for $U_{\rm SRG}(\alpha)$ is $U_{\rm SRG}(0) = \openone$.
Alternative to $\alpha$, it is common to use $\lambda_{\rm SRG} = \alpha^{-1/4}$
 for controlling the flow equation Eq.~(\ref{eq:uevo}).
The Hamiltonian is transformed by Eq.~(\ref{eq:h_srg}) from
 $\lambda_{\rm SRG} = \infty$ fm$^{-1}$
 to lower values where the interaction is "soft" enough for
 convergence of the many-body calculation methods.
As discussed, for example in Ref.~\cite{Jurgenson2009a},
 the SRG evolution, however, induces the many-body forces:
 \begin{equation}
   H(\lambda_{\rm SRG}) = T + V^{NN}(\lambda_{\rm SRG}) + V^{3N}(\lambda_{\rm SRG}) + \cdots.
 \end{equation}
Consequently, during the SRG evolution, we should keep many-body terms, as many as possible,
 even if the starting Hamiltonian does not include the many-body interactions.
In this work, three types of Hamiltonians are used.
First one, labeled by "$NN$--only", is obtained by keeping only
 the $NN$ interaction during the SRG evolution starting without the genuine
 $3N$ interaction.
Second one "$NN+3N$--ind" is obtained by keeping
 the $NN$ and $3N$ interactions during the SRG evolution starting without the genuine
 $3N$ interaction.
Third one "$NN+3N$--full" is obtained by keeping
 the $NN$ and $3N$ interactions during the SRG evolution starting with the genuine
 $3N$ interaction.

To evaluate nuclear root-mean-squared radii, we should transform the
 radius operator in the same manner as the Hamiltonian:
\begin{equation}
  \label{eq:revo}
  r^{2}(\alpha) = U^{\dag}_{\rm SRG}(\alpha) r^{2} U_{\rm SRG}(\alpha).
\end{equation}
The original radius operator is defined as
\begin{equation}
  r^{2} = r^{2(2)} = \frac{1}{A^{2}} \sum_{i<j} ({\mathbf r}_{i} - {\mathbf r}_{j})^{2}
\end{equation}
 with the coordinate vector of the $i$--th nucleon ${\mathbf r}_{i}$ and
 number of nucleon $A$.
In the same manner as the Hamiltonian, the many-body radius operator can be induced through the SRG evolution:
\begin{equation}
  r^{2}(\lambda_{\rm SRG}) = r^{2(2)}(\lambda_{\rm SRG}) +
  r^{2(3)}(\lambda_{\rm SRG}) + \cdots.
\end{equation}
Following to Refs.~\cite{Schuster2014, Schuster2015}, we keep up to three-body terms.

To perform many-body calculations, it is numerically efficient to
 transform to the laboratory frame.
Then, our Hamiltonian and radius operators can be rewritten as
\begin{align}
  H &= T^{(1)} + [T^{(2)} + V^{NN}(\lambda_{\rm SRG})] + V^{3N}(\lambda_{\rm SRG}), \\
  r^{2} &= r^{2(1)} + [r^{2(2)}(\lambda_{\rm SRG}) - r^{2(1)}] + r^{2(3)}(\lambda_{\rm SRG}).
\end{align}
Here, we use $T^{(1)}=\sum_{i}(1 - 1/A){\mathbf p}_{i}^{2}/2m$ with the $i$--th nucleon momentum ${\mathbf p}_{i}$
 and nucleon mass $m$, $T^{(2)}=-\sum_{i<j}{\mathbf p}_{i} \cdot {\mathbf p}_{j}/mA$,
 and $r^{2(1)} = (1 - 1/A) \sum_{i}{\mathbf r}^{2}_{i}$.
Note that $r^{2(1)}$ is chosen so that $r^{2(2)}(\lambda_{\rm SRG}) - r^{2(1)}$
 goes $- \sum_{i<j} {\mathbf r}_{i} \cdot {\mathbf r}_{j}/A$ in the limit of
 $\lambda_{\rm SRG} \rightarrow \infty$.
In the second quantization form, they are
\begin{align}
  \label{H:2ndq}
  H &= \sum_{a_{1}a_{2}} t_{a_{1}a_{2}}c^{\dag}_{a_{1}}c_{a_{2}} \notag \\
    & + \left(\frac{1}{2!}\right)^{2}\sum_{a_{1}a_{2}a_{3}a_{4}}
  v^{(2)}_{a_{1}a_{2}a_{3}a_{4}} c^{\dag}_{a_{1}}c^{\dag}_{a_{2}}
  c_{a_{4}}c_{a_{3}} \notag \\
  & + \left(\frac{1}{3!}\right)^{2} \sum_{a_{1}a_{2}a_{3}a_{4}a_{5}a_{6}}
  v^{(3)}_{a_{1}a_{2}a_{3}a_{4}a_{5}a_{6}}
  c^{\dag}_{a_{1}}c^{\dag}_{a_{2}} c^{\dag}_{a_{3}}
  c_{a_{6}} c_{a_{5}} c_{a_{4}}, \\
  \label{r:2ndq}
  r^{2} &= \sum_{a_{1}a_{2}} r^{2(1)}_{a_{1}a_{2}}c^{\dag}_{a_{1}}c_{a_{2}} \notag \\
    &+ \left(\frac{1}{2!}\right)^{2}\sum_{a_{1}a_{2}a_{3}a_{4}}
  r^{2(2)}_{a_{1}a_{2}a_{3}a_{4}} c^{\dag}_{a_{1}}c^{\dag}_{a_{2}}
  c_{a_{4}}c_{a_{3}} \notag \\
  & + \left(\frac{1}{3!}\right)^{2} \sum_{a_{1}a_{2}a_{3}a_{4}a_{5}a_{6}}
  r^{2(3)}_{a_{1}a_{2}a_{3}a_{4}a_{5}a_{6}}
  c^{\dag}_{a_{1}}c^{\dag}_{a_{2}} c^{\dag}_{a_{3}}
  c_{a_{6}} c_{a_{5}} c_{a_{4}}.
 \end{align}
Here, $c_{a}$ ($c^{\dag}_{a}$) is the annihilation (creation) operator of
 the nucleon at the state $a$.
In Eqs.~(\ref{H:2ndq}) and (\ref{r:2ndq}), shorthand notations,
\begin{align}
  t_{a_{1}a_{2}} &= \bra a_{1}|T^{(1)}|a_{2} \ket, \\
  v^{(2)}_{a_{1}a_{2}a_{3}a_{4}} &= \bra a_{1}a_{2} | V^{NN}(\lambda_{\rm SRG}) + T^{(2)} | a_{3}a_{4} \ket,  \\
  v^{(3)}_{a_{1}a_{2}a_{3}a_{4}a_{5}a_{6}} &=
  \bra a_{1}a_{2}a_{3} | V^{3N}(\lambda_{\rm SRG}) | a_{4}a_{5}a_{6} \ket, \\
 r^{2(1)}_{a_{1}a_{2}} &= \bra a_{1}| r^{2(1)} | a_{2} \ket,  \\
  r^{2(2)}_{a_{1}a_{2}a_{3}a_{4}} &= \bra a_{1}a_{2} | r^{2(2)}(\lambda_{\rm SRG})
  - \frac{1}{A-1} r^{2(1)}| a_{3}a_{4} \ket,  \\
 r^{2(3)}_{a_{1}a_{2}a_{3}a_{4}a_{5}a_{6}} &=
  \bra a_{1}a_{2}a_{3} | r^{2(3)}(\lambda_{\rm SRG}) | a_{4}a_{5}a_{6} \ket,
 \end{align}
are used for one-body-kinetic-term, two-body-interaction, three-body-interaction,
 one-body-radius, two-body-radius, and three-body-radius matrix elements, respectively.
The factor $1/(A-1)$ in two-body-radius matrix element is due to the normalization when a one-body
 operator is used as a two-body operator.
Because of the computational limitation, however, the direct treatment of the three-body
 matrix elements is still challenging.
Therefore, we follow the recent nuclear {\it ab initio} studies and introduce the normal-ordered
 two-body (NO2B) approximation~\cite{Roth2012, Binder2013}.
The key of the approximation is a rearrangement of the three-body term with respect to
 a reference state $|\Phi\ket$.
After the rearrangement, the zero-, one-, two-, and three-body pieces show up.
In the NO2B approximation, the residual three-body piece is discarded.
To apply to the UMOA framework, we take normal order again with respect to the
 nucleon vacuum state.
Then, the Hamiltonian is
\begin{align}
  H &\approx h^{(0),{\rm NO2B}} + \sum_{a_{1}a_{2}} h^{(1),{\rm NO2B}}_{a_{1}a_{2}}c^{\dag}_{a_{1}}c_{a_{2}} \notag \\
     & + \left(\frac{1}{2!}\right)^{2}\sum_{a_{1}a_{2}a_{3}a_{4}}
  h^{(2),{\rm NO2B}}_{a_{1}a_{2}a_{3}a_{4}} c^{\dag}_{a_{1}}c^{\dag}_{a_{2}}
   c_{a_{4}}c_{a_{3}}
   \label{h:2ndq-no}
\end{align}
 with
\begin{align}
  h^{(0), {\rm NO2B}} &= \frac{1}{6} \sum_{abc} v^{(3)}_{abcabc}n_{a} n_{b}n_{c}, \\
  h^{(1), {\rm NO2B}}_{a_{1}a_{2}} &= t_{a_{1}a_{2}} - \frac{1}{2} \sum_{bc}
     v^{(3)}_{a_{1}bca_{2}bc} n_{b}n_{c}, \\
  h^{(2), {\rm NO2B}}_{a_{1}a_{2}a_{3}a_{4}} &=
   v^{(2)}_{a_{1}a_{2}a_{3}a_{4}} + \sum_{b} v^{(3)}_{a_{1}a_{2}ba_{3}a_{4}b} n_{b}.
\end{align}
Here, $n_{a}$ is an occupation number for the orbit $a$, i.e. $n_{a}=1$ ($n_{a} = 0$)
  where $a$ is below (above) the Fermi level.
To minimize the effect of the truncated residual three-body piece,
 the choice of $| \Phi \ket$ is crucial.
In this work, we use the Hartree-Fock state as $|\Phi \ket$.
Same as the Hamiltonian, we employ the NO2B approximated radius operator:
\begin{align}
  r^{2} &\approx r^{2(0), {\rm NO2B}} + \sum_{a_{1}a_{2}} r^{2(1), {\rm NO2B}}_{a_{1}a_{2}}c^{\dag}_{a_{1}}c_{a_{2}} \notag \\
     & + \left(\frac{1}{2!}\right)^{2}\sum_{a_{1}a_{2}a_{3}a_{4}}
  r^{2(2), {\rm NO2B}}_{a_{1}a_{2}a_{3}a_{4}} c^{\dag}_{a_{1}}c^{\dag}_{a_{2}}
   c_{a_{4}}c_{a_{3}}.
   \label{r:2ndq-no}
\end{align}
 with
\begin{align}
  r^{2(0), {\rm NO2B}} &= \frac{1}{6} \sum_{abc} r^{2(3)}_{abcabc}n_{a} n_{b}n_{c}, \\
  r^{2(1), {\rm NO2B}}_{a_{1}a_{2}} &= r^{2(1)}_{a_{1}a_{2}}- \frac{1}{2} \sum_{bc}
     r^{2(3)}_{a_{1}bca_{2}bc} n_{b}n_{c}, \\
  r^{2(2), {\rm NO2B}}_{a_{1}a_{2}a_{3}a_{4}} &=
   r^{2(2)}_{a_{1}a_{2}a_{3}a_{4}} + \sum_{b} r^{2(3)}_{a_{1}a_{2}ba_{3}a_{4}b} n_{b}.
\end{align}

\section{unitary-model-operator approach \label{sec:umoa}}
\begin{fmffile}{MBPT}
\begin{table}
  \caption{\label{tab:mbpt} Hugenholtz diagrams for the ground-state energy up to the third order.
   Note that the first order contributions are omitted.
   The cross and dot indicate the one- and two-body part of Hamiltonian, respectively.
   The diagram rules are same as in Ref.~\cite{Shavitt2009}.}
  \begin{ruledtabular}
  \begin{tabular}{cccc}
    \multicolumn{4}{c}{Second order} \\ \hline
    & & & \\
    \multicolumn{2}{c}{
      \begin{fmfgraph}(50,50)
        \fmfstraight
        \fmfset{arrow_len}{0.25cm}
        \fmfset{arrow_ang}{15}
        \fmftop{v1}
          \fmfright{h1,h2}
          \fmfbottom{v2}
          \fmf{fermion,left=0.5}{v1,v2}
          \fmf{fermion,left=0.5}{v2,v1}
          \fmf{dashes,dash_len=0.01cm}{h1,v2}
          \fmf{dashes,dash_len=0.01cm}{h2,v1}
          \fmfv{decor.shape=cross,decor.size=0.25cm}{h1,h2}
      \end{fmfgraph}}
    &
    \multicolumn{2}{c}{
      \begin{fmfgraph}(50,50)
        \fmfstraight
        \fmfset{arrow_len}{0.25cm}
        \fmfset{arrow_ang}{15}
        \fmftop{v1}
          \fmfbottom{v2}
          \fmf{fermion,left=0.5}{v1,v2}
          \fmf{fermion,left=0.8}{v1,v2}
          \fmf{fermion,left=0.5}{v2,v1}
          \fmf{fermion,left=0.8}{v2,v1}
          \fmfdot{v1,v2}
      \end{fmfgraph}}
    \\
    \multicolumn{2}{c}{$S_{1}$}& \multicolumn{2}{c}{$S_{2}$} \\ \hline
    \multicolumn{4}{c}{Third order} \\ \hline
    & & & \\
      \begin{fmfgraph}(50,50)
        \fmfstraight
        \fmfset{arrow_len}{0.25cm}
        \fmfset{arrow_ang}{15}
        \fmftop{v1}
          \fmfbottom{v3}
          \fmfright{h3,h2,h1}
          \fmf{fermion,tension=100}{v1,v2}
          \fmf{fermion,tension=100}{v2,v3}
          \fmf{fermion,left=0.5}{v3,v1}
          \fmf{dashes}{h1,v1}
          \fmf{dashes}{h2,v2}
          \fmf{dashes}{h3,v3}
          \fmfv{decor.shape=cross,decor.size=0.25cm}{h1,h2,h3}
      \end{fmfgraph}
      &
      \begin{fmfgraph}(50,50)
        \fmfstraight
        \fmfset{arrow_len}{0.25cm}
        \fmfset{arrow_ang}{15}
        \fmftop{v1}
          \fmfbottom{v3}
          \fmfright{h3,h2,h1}
          \fmf{fermion,tension=100}{v2,v1}
          \fmf{fermion,tension=100}{v3,v2}
          \fmf{fermion,right=0.5}{v1,v3}
          \fmf{dashes}{h1,v1}
          \fmf{dashes}{h2,v2}
          \fmf{dashes}{h3,v3}
          \fmfv{decor.shape=cross,decor.size=0.25cm}{h1,h2,h3}
      \end{fmfgraph}
      &
      \begin{fmfgraph}(50,50)
        \fmfstraight
        \fmfset{arrow_len}{0.25cm}
        \fmfset{arrow_ang}{15}
        \fmftop{v1}
          \fmfbottom{v3}
          \fmfright{h3,h2,h1}
          \fmf{phantom,tension=100}{v1,v2}
          \fmf{phantom,tension=100}{v2,v3}
          \fmf{fermion,left=0.5}{v3,v1}
          \fmf{fermion,left=0.5}{v1,v3}
          \fmf{fermion,left=0.5}{v3,v2}
          \fmf{fermion,left=0.5}{v2,v3}
          \fmfdot{v3}
          \fmf{dashes}{h1,v1}
          \fmf{dashes}{h2,v2}
          \fmfv{decor.shape=cross,decor.size=0.25cm}{h1,h2}
      \end{fmfgraph}
      &
      \begin{fmfgraph}(50,50)
        \fmfstraight
        \fmfset{arrow_len}{0.25cm}
        \fmfset{arrow_ang}{15}
        \fmftop{v1}
          \fmfbottom{v3}
          \fmfright{h3,h2,h1}
          \fmf{phantom,tension=100}{v1,v2}
          \fmf{phantom,tension=100}{v2,v3}
          \fmf{fermion,left=0.8}{v2,v1}
          \fmf{fermion,left=0.8}{v1,v2}
          \fmf{fermion,left=0.8}{v3,v2}
          \fmf{fermion,left=0.8}{v2,v3}
          \fmfdot{v2}
          \fmf{dashes}{h1,v1}
          \fmf{dashes}{h3,v3}
          \fmfv{decor.shape=cross,decor.size=0.25cm}{h1,h3}
      \end{fmfgraph}
      \\
      $T_{1}$ & $T_{2}$ & $T_{3}$ & $T_{4}$
      \\
      \\
      \begin{fmfgraph}(50,50)
        \fmfstraight
        \fmfset{arrow_len}{0.25cm}
        \fmfset{arrow_ang}{15}
        \fmftop{v1}
          \fmfbottom{v3}
          \fmfright{h3,h2,h1}
          \fmf{phantom,tension=100}{v1,v2}
          \fmf{phantom,tension=100}{v2,v3}
          \fmf{fermion,left=0.5}{v2,v1}
          \fmf{fermion,left=0.5}{v1,v2}
          \fmf{fermion,left=0.5}{v3,v1}
          \fmf{fermion,left=0.5}{v1,v3}
          \fmfdot{v1}
          \fmf{dashes}{h2,v2}
          \fmf{dashes}{h3,v3}
          \fmfv{decor.shape=cross,decor.size=0.25cm}{h2,h3}
      \end{fmfgraph}
      &
      \begin{fmfgraph}(50,50)
        \fmfstraight
        \fmfset{arrow_len}{0.25cm}
        \fmfset{arrow_ang}{15}
        \fmftop{v1}
          \fmfbottom{v3}
          \fmfright{h3,h2,h1}
          \fmf{fermion,left=0.8}{v3,v1}
          \fmf{fermion,tension=100}{v1,v2}
          \fmf{fermion,tension=100}{v2,v3}
          \fmf{fermion,left=0.5}{v3,v2}
          \fmf{fermion,left=0.5}{v2,v3}
          \fmfdot{v2,v3}
          \fmf{dashes}{h1,v1}
          \fmfv{decor.shape=cross,decor.size=0.25cm}{h1}
      \end{fmfgraph}
      &
      \begin{fmfgraph}(50,50)
        \fmfstraight
        \fmfset{arrow_len}{0.25cm}
        \fmfset{arrow_ang}{15}
        \fmftop{v1}
          \fmfbottom{v3}
          \fmfright{h3,h2,h1}
          \fmf{fermion,right=0.8}{v1,v3}
          \fmf{fermion,tension=100}{v2,v1}
          \fmf{fermion,tension=100}{v3,v2}
          \fmf{fermion,left=0.5}{v3,v2}
          \fmf{fermion,left=0.5}{v2,v3}
          \fmfdot{v2,v3}
          \fmf{dashes}{h1,v1}
          \fmfv{decor.shape=cross,decor.size=0.25cm}{h1}
      \end{fmfgraph}
      &
      \begin{fmfgraph}(50,50)
        \fmfstraight
        \fmfset{arrow_len}{0.25cm}
        \fmfset{arrow_ang}{15}
        \fmftop{v1}
          \fmfbottom{v3}
          \fmfright{h3,h2,h1}
          \fmf{fermion,right=0.8}{v1,v3}
          \fmf{fermion,tension=100}{v2,v1}
          \fmf{fermion,tension=100}{v3,v2}
          \fmf{fermion,left=0.5}{v3,v1}
          \fmf{fermion,left=0.5}{v1,v3}
          \fmfdot{v1,v3}
          \fmf{dashes}{h2,v2}
          \fmfv{decor.shape=cross,decor.size=0.25cm}{h2}
      \end{fmfgraph}
      \\
      $T_{5}$ & $T_{6}$ & $T_{7}$ & $T_{8}$
      \\
      \\
      \begin{fmfgraph}(50,50)
        \fmfstraight
        \fmfset{arrow_len}{0.25cm}
        \fmfset{arrow_ang}{15}
        \fmftop{v1}
          \fmfbottom{v3}
          \fmfright{h3,h2,h1}
          \fmf{fermion,left=0.8}{v3,v1}
          \fmf{fermion,tension=100}{v1,v2}
          \fmf{fermion,tension=100}{v2,v3}
          \fmf{fermion,left=0.5}{v3,v1}
          \fmf{fermion,left=0.5}{v1,v3}
          \fmfdot{v1,v3}
          \fmf{dashes}{h2,v2}
          \fmfv{decor.shape=cross,decor.size=0.25cm}{h2}
      \end{fmfgraph}
      &
      \begin{fmfgraph}(50,50)
        \fmfstraight
        \fmfset{arrow_len}{0.25cm}
        \fmfset{arrow_ang}{15}
        \fmftop{v1}
          \fmfbottom{v3}
          \fmfright{h3,h2,h1}
          \fmf{fermion,left=0.8}{v3,v1}
          \fmf{fermion,tension=100}{v1,v2}
          \fmf{fermion,tension=100}{v2,v3}
          \fmf{fermion,left=0.5}{v2,v1}
          \fmf{fermion,left=0.5}{v1,v2}
          \fmfdot{v1,v2}
          \fmf{dashes}{h3,v3}
          \fmfv{decor.shape=cross,decor.size=0.25cm}{h3}
      \end{fmfgraph}
      &
      \begin{fmfgraph}(50,50)
        \fmfstraight
        \fmfset{arrow_len}{0.25cm}
        \fmfset{arrow_ang}{15}
        \fmftop{v1}
          \fmfbottom{v3}
          \fmfright{h3,h2,h1}
          \fmf{fermion,right=0.8}{v1,v3}
          \fmf{fermion,tension=100}{v2,v1}
          \fmf{fermion,tension=100}{v3,v2}
          \fmf{fermion,left=0.5}{v2,v1}
          \fmf{fermion,left=0.5}{v1,v2}
          \fmfdot{v1,v2}
          \fmf{dashes}{h3,v3}
          \fmfv{decor.shape=cross,decor.size=0.25cm}{h3}
      \end{fmfgraph}
      &
      \begin{fmfgraph}(50,50) 
        \fmfstraight
        \fmfset{arrow_len}{0.25cm}
        \fmfset{arrow_ang}{15}
        \fmftop{v1}
          \fmfbottom{v3}
          \fmf{fermion,left=0.8}{v1,v3}
          \fmf{fermion,right=0.8}{v1,v3}
          \fmf{fermion,left=0.5}{v2,v1}
          \fmf{fermion,right=0.5}{v2,v1}
          \fmf{fermion,left=0.5}{v3,v2}
          \fmf{fermion,right=0.5}{v3,v2}
          \fmfdot{v1,v2,v3}
      \end{fmfgraph}
      \\
      $T_{9}$ & $T_{10}$ & $T_{11}$ & $T_{12}$ \\
      \\
      \begin{fmfgraph}(50,50) 
        \fmfstraight
        \fmfset{arrow_len}{0.25cm}
        \fmfset{arrow_ang}{15}
        \fmftop{v1}
          \fmfbottom{v3}
          \fmf{fermion,left=0.8}{v3,v1}
          \fmf{fermion,right=0.8}{v3,v1}
          \fmf{fermion,left=0.5}{v1,v2}
          \fmf{fermion,right=0.5}{v1,v2}
          \fmf{fermion,left=0.5}{v2,v3}
          \fmf{fermion,right=0.5}{v2,v3}
          \fmfdot{v1,v2,v3}
      \end{fmfgraph}
      &
      \begin{fmfgraph}(50,50) 
        \fmfstraight
        \fmfset{arrow_len}{0.25cm}
        \fmfset{arrow_ang}{15}
        \fmftop{v1}
          \fmfbottom{v3}
          \fmf{fermion,left=0.8}{v3,v1}
          \fmf{fermion,left=0.8}{v1,v3}
          \fmf{fermion,left=0.5}{v2,v1}
          \fmf{fermion,left=0.5}{v1,v2}
          \fmf{fermion,left=0.5}{v3,v2}
          \fmf{fermion,left=0.5}{v2,v3}
          \fmfdot{v1,v2,v3}
      \end{fmfgraph}
      & & \\
      $T_{13}$ & $T_{14}$ &  & \\
  \end{tabular}
  \end{ruledtabular}
\end{table}
\end{fmffile}
To solve the many-body Schr\"odinger equation associated with the Hamiltonian Eq.~(\ref{h:2ndq-no}),
 the UMOA~\cite{Suzuki1987, Suzuki1988, Suzuki1994, Miyagi2017a} is employed in this work.
In the UMOA, we construct the effective Hamiltonian with the unitary transformation:
\begin{equation}
  \label{eq:utr-h}
  \widetilde{H} = U^{\dag} H U.
\end{equation}
The $U$ is defined by the product of two exponential operators,
\begin{equation}
U = e^{S^{(1)}}e^{S^{(2)}},
\end{equation}
where $S^{(1)}$ and $S^{(2)}$ are anti-hermitian one- and two-body correlation operators, respectively.
Note that the sole use of $S^{(1)}$ ($S^{(2)}=0$) reduces the UMOA to the Hartree-Fock (HF) theory.
The $S^{(1)}$ and $S^{(2)}$ are specified by applying iteratively the Okubo-Lee-Suzuki
 method~\cite{Okubo1954, Lee1980, Suzuki1980} so that $\widetilde{H}$
 does not induce the one-particle-one-hole and two-particle-two-hole
 excitations from the reference state $|\Phi\ket$.
Since the unitary transformation~(\ref{eq:utr-h}) can induce many-body interactions,
 $\widetilde{H}$ can include many-body operators even if the original Hamiltonian
 is restricted up to the two-body interaction.
In actual calculations, we decompose $\widetilde{H}$ with the cluster expansion and
 truncate the effect of the four- and higher-body cluster terms:
\begin{align}
  \widetilde{H} &\approx
  \sum_{a_{1}a_{2}} \widetilde{H}^{(1)}_{a_{1}a_{2}}c^{\dag}_{a_{1}}c_{a_{2}}  \notag \\
  & +
  \frac{1}{4} \sum_{a_{1}a_{2}a_{3}a_{4}} \widetilde{H}^{(2)}_{a_{1}a_{2}a_{3}a_{4}}
  c^{\dag}_{a_{1}} c^{\dag}_{a_{2}} c_{a_{4}} c_{a_{3}} \notag \\
  & +
  \frac{1}{36} \sum_{a_{1}a_{2}a_{3}a_{4}a_{5}a_{6}}
  \widetilde{H}^{(3)}_{a_{1}a_{2}a_{3}a_{4}a_{5}a_{6}}c^{\dag}_{a_{1}} c^{\dag}_{a_{2}}
  c^{\dag}_{a_{3}} c_{a_{6}} c_{a_{5}} c_{a_{4}},
\end{align}
where $\widetilde{H}^{(1)}_{a_{1}a_{2}}$, $\widetilde{H}^{(2)}_{a_{1}a_{2}a_{3}a_{4}}$,
 and $\widetilde{H}^{(3)}_{a_{1}a_{2}a_{3}a_{4}a_{5}a_{6}}$ are the one-, two-, and three-body
 matrix elements, respectively (see, for example, Ref.~\cite{Miyagi2017a} for more details).
Then, the ground-state energy $E_{\rm g.s.}$ can be obtained approximately by
\begin{align}
  \label{eq:egs}
  E_{\rm g.s.} &\approx E^{\rm 1,2BC} + E^{\rm 3BC}, \\
  E^{\rm 1,2BC} &= \sum_{a}\widetilde{H}^{(1)}_{aa}n_{a} + \frac{1}{2}
  \sum_{ab}\widetilde{H}^{(2)}_{abab}n_{a}n_{b}, \\
  E^{\rm 3BC} &=
   \frac{1}{6} \sum_{abc}\widetilde{H}^{(3)}_{abcabc}n_{a}n_{b}n_{c}.
\end{align}
Since the direct treatment of three-body term demands huge computational resources, however,
 the contribution of three-body cluster term is approximately evaluated up to second order of $S^{(2)}$~\cite{Suzuki1994}:
 \begin{align}
   & E^{\rm 3BC} \approx
    \frac{1}{4} \sum_{abcd} \sum_{ef}
    \widetilde{H}^{(2)}_{abcd} S^{(2)}_{efab} S^{(2)}_{efcd} n_{a}n_{b}n_{c}n_{d}
    \bar{n}_{e}\bar{n}_{f} \notag \\
    \label{eq:e3bc}
    & \hspace{2em}
    + \sum_{abc} \sum_{def}
    \widetilde{H}^{(2)}_{adcf} S^{(2)}_{debc} S^{(2)}_{efab} n_{a}n_{b}n_{c}\bar{n}_{d}
    \bar{n}_{e}\bar{n}_{f}.
 \end{align}
The $S^{(2)}_{abcd}$ is the matrix element of the two-body correlation operator and
 $\bar{n}_{a} = 1 - n_{a}$ is used.
To clarify the contribution of each cluster term, the
 comparison with the many-body perturbation theory (MBPT) would be useful.
Table~\ref{tab:mbpt} shows the diagrams for the ground-state energy from the third-order MBPT.
Following to the perturbative derivation of correlation operators, shown for example in Refs~\cite{Suzuki1983,Suzuki1984},
 the contribution of each cluster term to the ground-state energy can be derived.
In terms of the many-body perturbation theory, $E^{\rm 1,2BC}$, $E^{\rm 3BC}$, and
 $E_{\rm g.s.}$ are
\begin{align}
  E^{\rm 1,2BC} &= E_{1} + \sum_{i=1}^{2}S_{i} + \sum_{i=1}^{12}T_{i} - T_{13} \notag \\
  & \hspace{4em} + (\text{higher order terms}), \\
  E^{\rm 3BC} &= 2 T_{13} + T_{14} + (\text{higher order terms}), \\
  E_{\rm g.s.} &= E_{1} + \sum_{i}^{2}S_{i} + \sum_{i=1}^{14}T_{i} \notag \\
  & \hspace{4em} + (\text{higher order terms}).
\end{align}
Here, $S_{i}$ and $T_{i}$ are the second- and third-order contributions shown in
Table~\ref{tab:mbpt}, respectively, and $E_{1}$ is the first order ground-state energy.
At one-plus-two-body cluster level, the third-order diagrams are not completed.
The three-body cluster term contributions compensate the third order~\cite{Suzuki1986a}.
Note that $S_{1}$ and $T_{1}$ to $T_{11}$ vanishes when the HF basis is employed.

To evaluate the expectation value of the radius operator obtained in Eq.~(\ref{r:2ndq-no}),
 the effective operator $\widetilde{r}^{2}$ is used:
\begin{equation}
  \widetilde{r}^{2} = U^{\dag} r^{2} U.
\end{equation}
Similarly to the Hamiltonian, the unitary transformation of the radius operator induces the many-body terms.
However, results examined here are calculated keeping up to two-body terms and does not
 include any contributions from three- and higher-body terms~\cite{Miyagi2017a}:
\begin{align}
  \widetilde{r}^{2} & \approx \sum_{a_{1}a_{2}} \widetilde{r}^{2(1)}_{a_{1}a_{2}}
  c^{\dag}_{a_{1}} c_{a_{2}} \notag \\
  & + \frac{1}{4} \sum_{a_{1}a_{2}a_{3}a_{4}} \widetilde{r}^{2(2)}_{a_{1}a_{2}a_{3}a_{4}}
  c^{\dag}_{a_{1}} c^{\dag}_{a_{2}} c_{a_{4}} c_{a_{3}}.
\end{align}
Then, the mean-squared radius $r_{\rm g.s.}^{2}$ is approximately evaluated as
\begin{equation}
  r^{2}_{\rm g.s.} \approx
  \sum_{a}\widetilde{r}^{2(1)}_{aa} n_{a} +
  \frac{1}{2} \sum_{ab} \widetilde{r}^{2(2)}_{abab}
  n_{a}n_{b}.
\end{equation}

\section{Results and discussions \label{sec:res}}
\begin{table*}[]
  \caption{\label{tab1} Ground-state energies for ${}^{4}$He, ${}^{16}$O, and ${}^{40}$Ca. All the
    calculation results are obtained at $e_{\max}  = 14$ and $\hbar\omega = 25$ MeV.}
  \begin{ruledtabular}
  \begin{tabular}{clrrrr}
   & & \multicolumn{4}{c}{$E_{\rm g.s.}$ (MeV)} \\ \cline{3-6}
   Nuclide & $\lambda_{\rm SRG}$ (fm$^{-1}$) & $NN$--only & $NN+3N$--ind & $NN+3N$--full  & Exp.\cite{Wang2012} \\ \hline
           & 1.88             &$-27.94 $  & $-25.19$     &$ -27.81$   &          \\
 ${}^{4}$He& 2.0              &$-27.73 $  & $-25.18$     &$ -27.76$   &$-28.30$  \\
           & 2.24             &$-27.23 $  & $-25.16$     &$ -27.62$   &          \\ \\
           & 1.88             &$-167.79$  & $-119.33$    &$-127.16$   &          \\
 ${}^{16}$O& 2.0              &$-162.69$  & $-119.51$    &$-126.33$   &$-127.62$ \\
           & 2.24             &$-152.88$  & $-119.56$    &$-124.50$   &          \\ \\
           & 1.88             &$-615.62$  & $-349.08$    &$-368.44$   &          \\
${}^{40}$Ca& 2.0              &$-588.45$  & $-352.03$    &$-366.14$   &$-342.05$ \\
           & 2.24             &$-536.26$  & $-355.61$    &$-360.23$   &          \\
 \end{tabular}
  \end{ruledtabular}
\end{table*}

In this work, we use the next-to-next-to-next-to leading order
 (N$^{3}$LO) $NN$ interaction by Entem and Machleidt~\cite{Entem2003} and
 local form N$^{2}$LO $3N$ interaction~\cite{Navratil2007a} from $\chi$EFT.
Both two- and three-body SRG evolutions are done in the harmonic-oscillator (HO) space.
The two-body interactions are obtained from the $N_{\max} = 200$ space calculations.
Here, $N_{\max}$ is the boundary of the HO quantum number for the two-body relative coordinate and
 is $N_{\max} = \max(2n + l)$ with the radial quantum number $n$ and angular momentum $l$.
Following Ref.~\cite{Roth2014}, the three-body SRG evolution is done in ramp A model space defined
 in Fig.~3 in Ref~\cite{Roth2014}.
To obtain the three-body matrix element,
 the frequency conversion technique~\cite{Roth2014} is used with the
 parent HO energy $\hbar\omega = 35$ MeV matrix elements.
For N$^{2}$LO $3N$ interaction, we use $c_{D} = -0.2$, $c_{E} = 0.098$, and $\Lambda_{3N} = 400$
 MeV$/c$~\cite{Roth2012},
 so as to compare with the other {\it ab initio} calculation results.
Note that the low-energy constant $c_{D}$ used here does not fit
 the $^{3}$H half-life as claimed in the past~\cite{Gazit2009,Marcucci2012}.
The impact of the modification of the $3N$ force with the $c_{D}$ that
 fits $^{3}$H half-life will be discussed in the forthcoming publications.
The size of the contributions from induced many-body forces can be estimated from
 the SRG resolution scale, $\lambda_{\rm SRG}$, dependence of calculated results.
To do so, we employ three SRG resolution scales $\lambda_{\rm SRG} = 1.88$, $2.0$, and $2.24$ fm$^{-1}$.
The NO2B approximated Hamiltonian is obtained through the HF calculations at $e_{3\max}=14$.
Here, $e_{3\max}$ is introduced to handle the three-body matrix element and
 is $e_{3\max} = \max(2n_{1} + l_{1} + 2n_{2} + l_{2} + 2n_{3} + l_{3})$ with
 the single-particle radial quantum number $n_{i}$ $(i = 1, 2, 3)$ and
 angular momentum $l_{i}$ $(i = 1, 2, 3)$.
We checked that the changing from $e_{3\max} = 12$ to $e_{3\max} = 14$ affects less than
 1\% of total ground-state energies for nuclei calculated in the present work.
UMOA calculations are done in the model space defined
 by $e_{\max} = \max(2n_{1}+l_{1})$~\cite{Miyagi2017a}.

 \subsection{Ground-state energies}
\begin{figure}[t]
  \includegraphics[clip,width=\columnwidth]{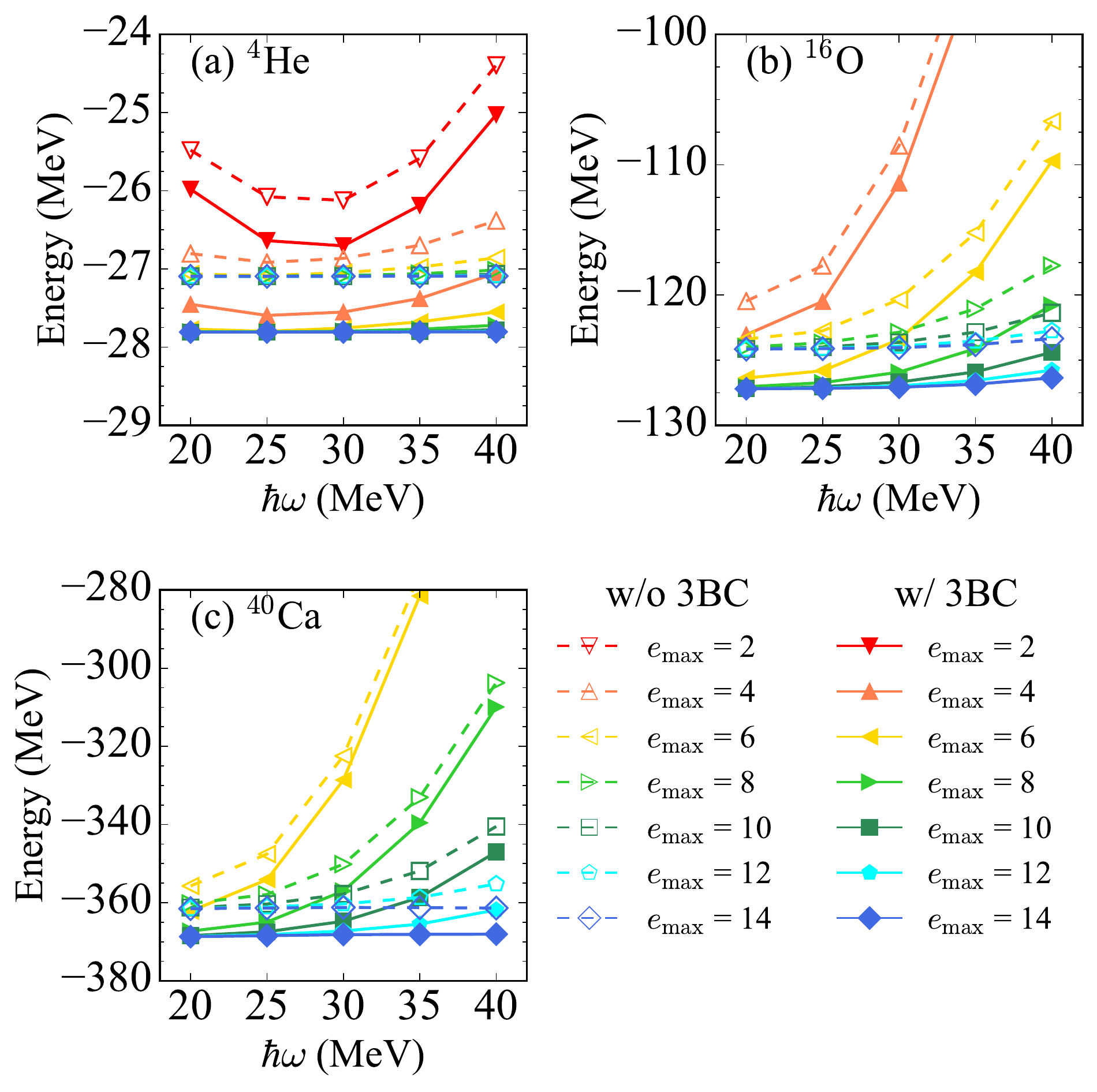}
  \caption{\label{fig:1} (color online) Ground-state energies for ${}^{4}$He, ${}^{16}$O, and
  ${}^{40}$Ca as functions of $\hbar\omega$ with the $NN+3N$--full interaction.
  The dashed (solid) lines calculated without (with) the three-body cluster term energy.
  The interaction is obtained by SRG evolution of chiral N$^{3}$LO $NN$~\cite{Entem2003} and N$^{2}$LO $3N$~\cite{Navratil2007, Roth2012}
  interactions up to $\lambda_{\rm SRG} = 1.88$ fm$^{-1}$.}
\end{figure}
Figure~\ref{fig:1} shows the convergence property of the ground-state
 energies for ${}^{4}$He, ${}^{16}$O, and ${}^{40}$Ca
 calculated with the $NN+3N$--full interaction from $\chi$EFT
 evolved up to $\lambda_{\rm SRG} = 1.88$ fm$^{-1}$.
Our calculations are done with varying $\hbar\omega$ and $e_{\max}$ to see the numerical convergence.
Note that the final results should not depend on $\hbar\omega$
 because the initial Hamiltonian Eq.~(\ref{eq:hini}) does not include $\hbar\omega$.
Similar to other {\it ab initio} calculations,
 our ground-state energies show parabolic $\hbar\omega$-dependence at small $e_{\max}$
 and gain with increasing $e_{\max}$.
For all cases examined here, $\hbar\omega$- and $e_{\max}$-independent
 results are obtained $e_{\max}=14$.
The results with $e_{\max}=14$ and $\hbar\omega=25$ MeV are used in the following discussion.

\begin{figure}[t]
  \includegraphics[width=\columnwidth, clip]{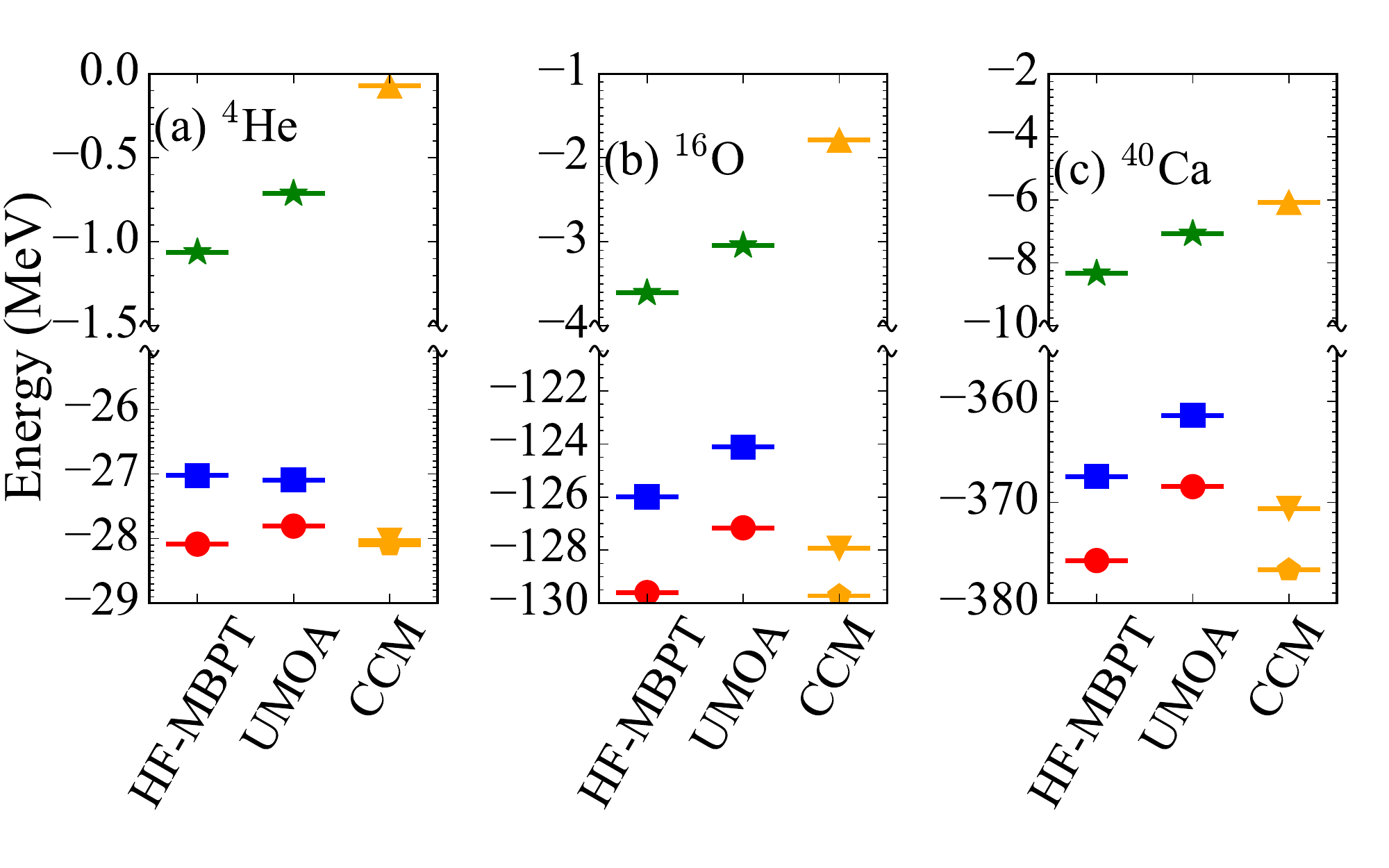}
  \caption{\label{fig:umoa-mbpt} (color online) Ground-state energies from the
 Hartree-Fock basis many-body perturbation theory (HF-MBPT), UMOA, and coupled-cluster method (CCM).
 The energies of HF-MBPT are calculated with Eqs.~(\ref{eq:hf-mbpt1})-(\ref{eq:hf-mbpt3}).
 The CCM results are taken from Refs.~\cite{Binder2014a,Binder2013a}.
 The interaction is obtained by SRG evolution of chiral N$^{3}$LO $NN$~\cite{Entem2003} and N$^{2}$LO $3N$~\cite{Navratil2007, Roth2012}
  interactions up to $\lambda_{\rm SRG} = 1.88$ fm$^{-1}$.
  For the HF-MBPT and UMOA energies, $E^{\rm 1,2BC}$ (square), $E^{\rm 3BC}$ (star), and $E_{\rm g.s.}$ (circle) are shown.
  For the CCM energies, CCSD (down triangle), triple correction (up triangle), and CR-CC(2,3) energies
  (pentagon) are shown.}
\end{figure}
To investigate the contributions of the cluster expansion, in Fig.~\ref{fig:umoa-mbpt},
 it is illustrated that the comparison among UMOA,
 Hartree-Fock basis many-body perturbation theory (HF-MBPT), and coupled-cluster method (CCM) energies.
In terms of HF-MBPT, the energies
 $E^{\rm 1,2BC}_{\rm MBPT}$, $E^{\rm 3BC}_{\rm MBPT}$, and $E_{\rm g.s., MBPT}$ are
 evaluated as
 \begin{align}
   \label{eq:hf-mbpt1}
   E^{\rm 1,2BC}_{\rm MBPT} &= E_{\rm HF} + S_{2} + T_{12} - T_{13}, \\
   \label{eq:hf-mbpt2}
   E^{\rm 3BC}_{\rm MBPT} &= 2T_{13} + T_{14}, \\
   \label{eq:hf-mbpt3}
   E_{\rm g.s., MBPT} &= E_{\rm HF} + S_{2} + T_{12} + T_{13} + T_{14},
 \end{align}
 with the Hartree-Fock energy $E_{\rm HF}$.
Note that $E_{\rm g.s., MBPT}$ is the third-order HF-MBPT energy.
In the figure, the UMOA and HF-MBPT energies are reasonably
 close to each other and it can be seen that
 the main contributions of $E^{\rm 3BC}$
 are from the third order hole-hole ($T_{13}$) and particle-hole ($T_{14}$)
 ladder diagrams.
Also, it is shown that the sum of the higher order terms taken into account in the UMOA is repulsive.
Comparing to CCM energies, total UMOA energies (circle) look closer to the CCSD energies (down triangle)
 than to the CR-CC(2,3) energies (pentagon).
The $E^{\rm 3BC}$ are
 $-0.71$, $-3.04$, and $-7.07$ MeV for ${}^{4}$He, ${}^{16}$O, and ${}^{40}$Ca,
 respectively, and are only a few percent of the total energies.
Since the contributions from four- and higher-body cluster terms are expected to be smaller than
 those from the three-body cluster term, the UMOA results are converged with respect to the cluster expansion.
For ${}^{16}$O, our ground-state energy $-127.16$ MeV is slightly underbound
 compared to the experimental energy ($-127.62$ MeV),
 while the recent {\em ab initio} calculation results show milidly
 overbound to the experiment, for
 example, $-130.6(1)$ MeV from in-medium SRG approach~\cite{Hergert2013}
 and $-129.7$ MeV from CCM~\cite{Binder2014a}.
Again, this disagreement between our and other {\it ab initio} results
 is same order of the size of the perturbative three-body-cluster contribution
 and consistent with the accuracy of the UMOA calculations.

\begin{figure}[t]
  \includegraphics[width=\columnwidth]{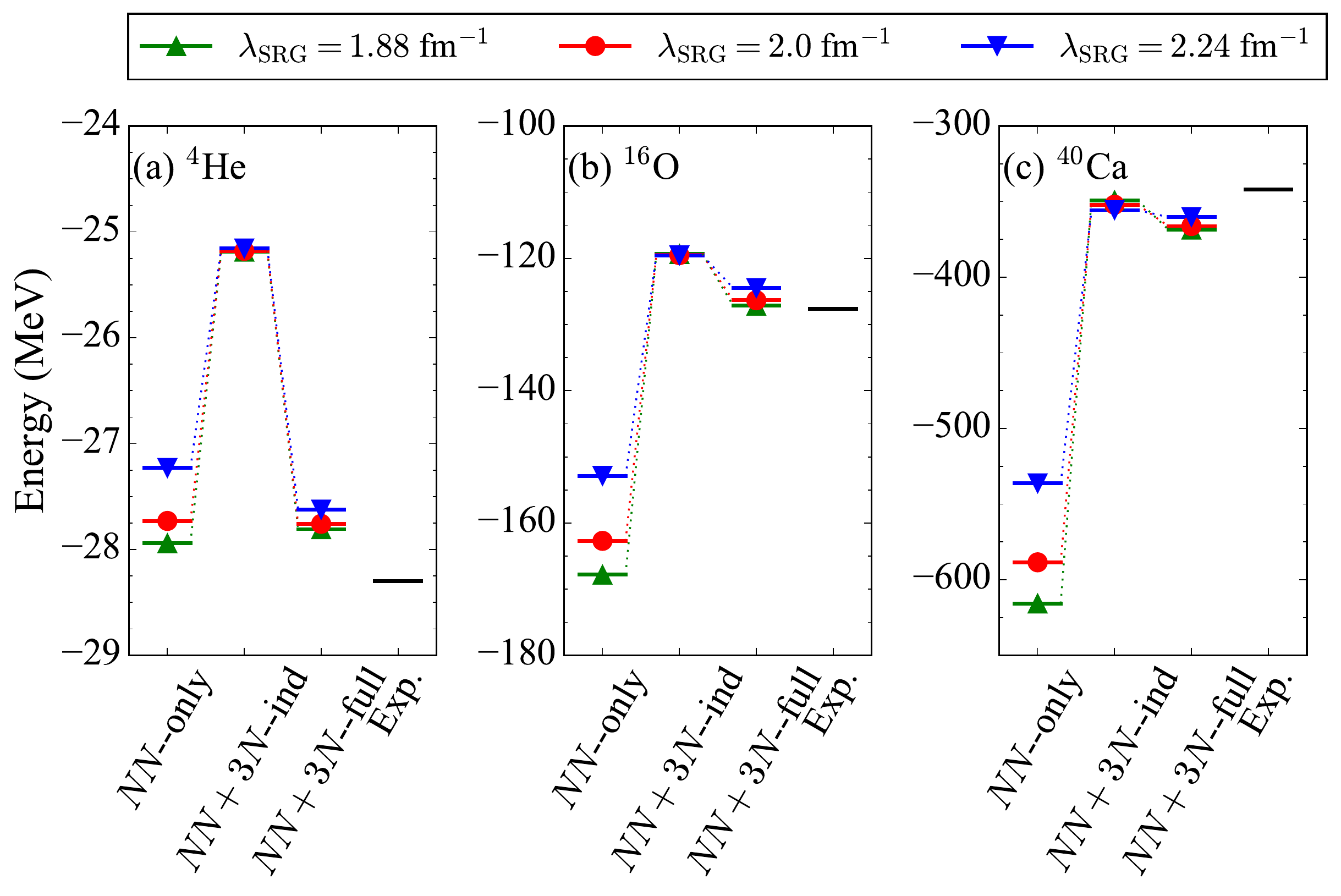}
  \caption{\label{fig:2} (color online) Ground-state energies for ${}^{4}$He, ${}^{16}$O, and
    ${}^{40}$Ca. All the calculation results are obtained at $e_{\max} = 14$ and
    $\hbar\omega = 25$ MeV.
    The experimental data are taken from Ref.~\cite{Wang2012}}
\end{figure}

As for calculations with $NN$--only and $NN+3N$--ind interactions,
 we observe the similar convergence pattern
 and find the converged results at $e_{\max} = 14$ calculations.
In Figure~\ref{fig:2}, the calculated ground-state energies are summarized together with
 the comparisons to the experimental data.
In case of $NN$--only interaction results, as the mass number increases,
 the ground-state energies show overbinding and $\lambda_{\rm SRG}$-dependence
 becomes considerable.
By taking the SRG induced $3N$ interaction into account, the $\lambda_{\rm SRG}$-dependence
 is drastically reduced and ground-state energies rise.
This $\lambda_{\rm SRG}$-independence of ground-state energies
 implies that the contributions from SRG induced four- and many-body
 interactions are negligible.
With the genuine $\chi$EFT N${}^{2}$LO $3N$ interaction, the calculated ground-state energies
 are comparable to the experimental data for ${}^{4}$He and ${}^{16}$O, while
 overbinding is seen for ${}^{40}$Ca.
The current choice of the genuine $3N$ interaction gives 9\%, 6\%, and 4\% attractions
 for ${}^{4}$He, ${}^{16}$O, and ${}^{40}$Ca, respectively.
The energies presented in Fig.~\ref{fig:2} are also displayed in Table~\ref{tab1}.
Our ground-state energies show reasonable agreement with the other {\it ab initio} results
 from the same interaction~\cite{Hergert2013, Roth2012, Tichai2016, Cipollone2013, Soma2014,
 Binder2013}.
The explicit treatment of the three-body cluster term seems to be necessary to discuss more
 precisely the accuracy of the UMOA calculation.
Such works are on going and will be reported in the future publication.

 \subsection{Root-mean square radii}
\begin{figure}[t]
  \includegraphics[width=\columnwidth, clip]{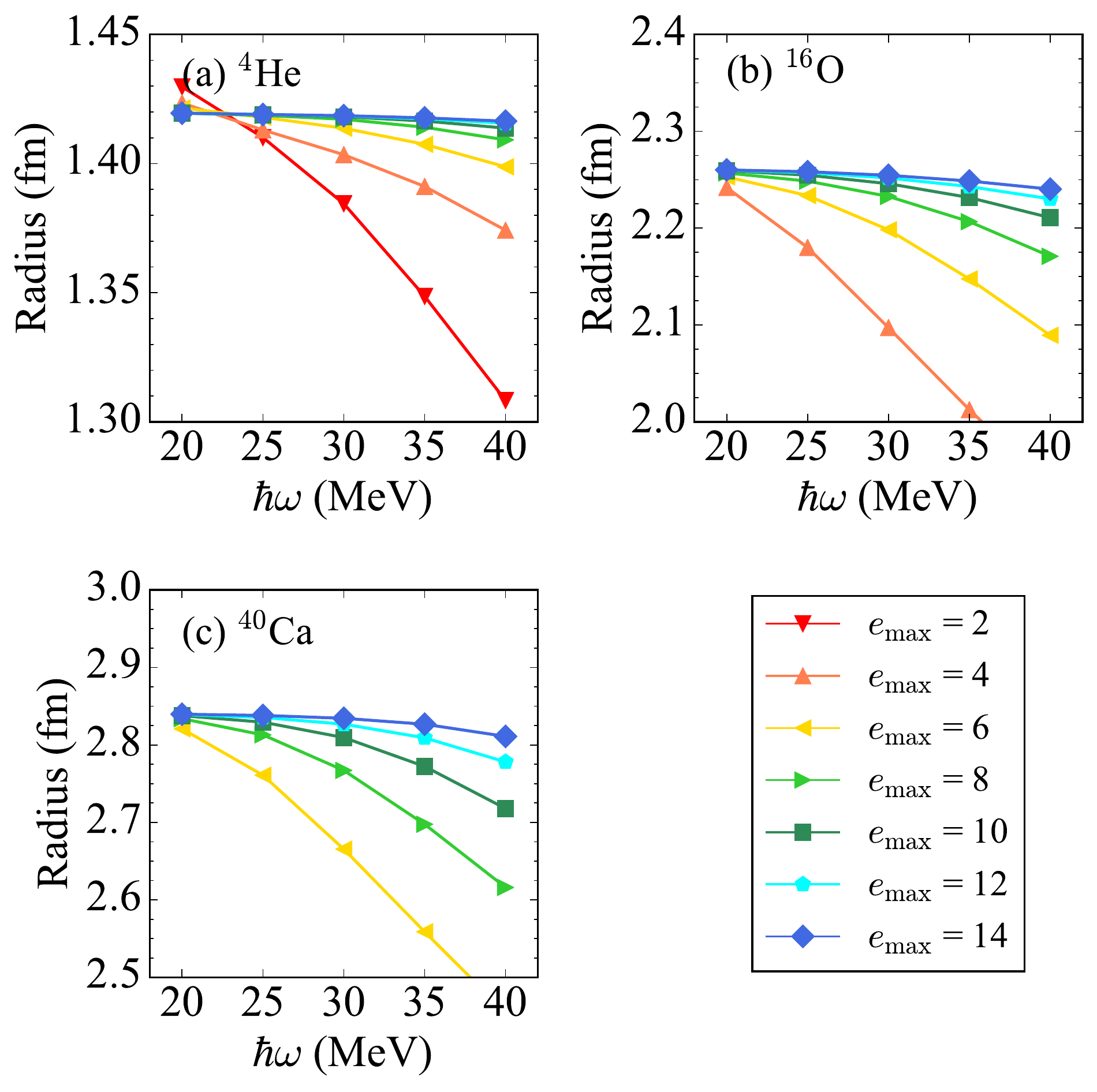}
  \caption{\label{fig:3} (color online) Expectation values of bare root-mean-squared radius operator for ${}^{4}$He, ${}^{16}$O, and
  ${}^{40}$Ca as functions of $\hbar\omega$. Here, $NN+3N$--full interaction
  at $\lambda_{\rm SRG} = 1.88$ fm$^{-1}$ is employed.}
\end{figure}

In the same as the ground-state energy calculations, we calculate the expectation values of the
 bare root-mean-squared radius operator with the chiral
 $NN+3N$--full interaction at $\lambda_{\rm SRG} = 1.88$ fm ${}^{-1}$
 varying both of $\hbar\omega$ and $e_{\max}$ to examine the
 convergence.
The results for ${}^{4}$He, ${}^{16}$O, and ${}^{40}$Ca are illustrated in Fig.~\ref{fig:3}.
As demonstrated in the figure, calculated radii become $\hbar\omega$- and $e_{\max}$-independent with
 increasing $e_{\max}$.
At $\hbar\omega = 25$ MeV, we find the converged radii within $0.01$ fm for all nuclei calculated here.
Note that our converged radius of $2.84$ fm for ${}^{40}$Ca from the interaction evolved up
 to $\lambda_{\rm SRG}=2.0$
 fm$^{-1}$ shows reasonable agreement with
 the SCGF result of $2.89$ fm~\cite{Soma2014} from the same interaction.

\begin{figure}[t]
  \includegraphics[width=\columnwidth, clip]{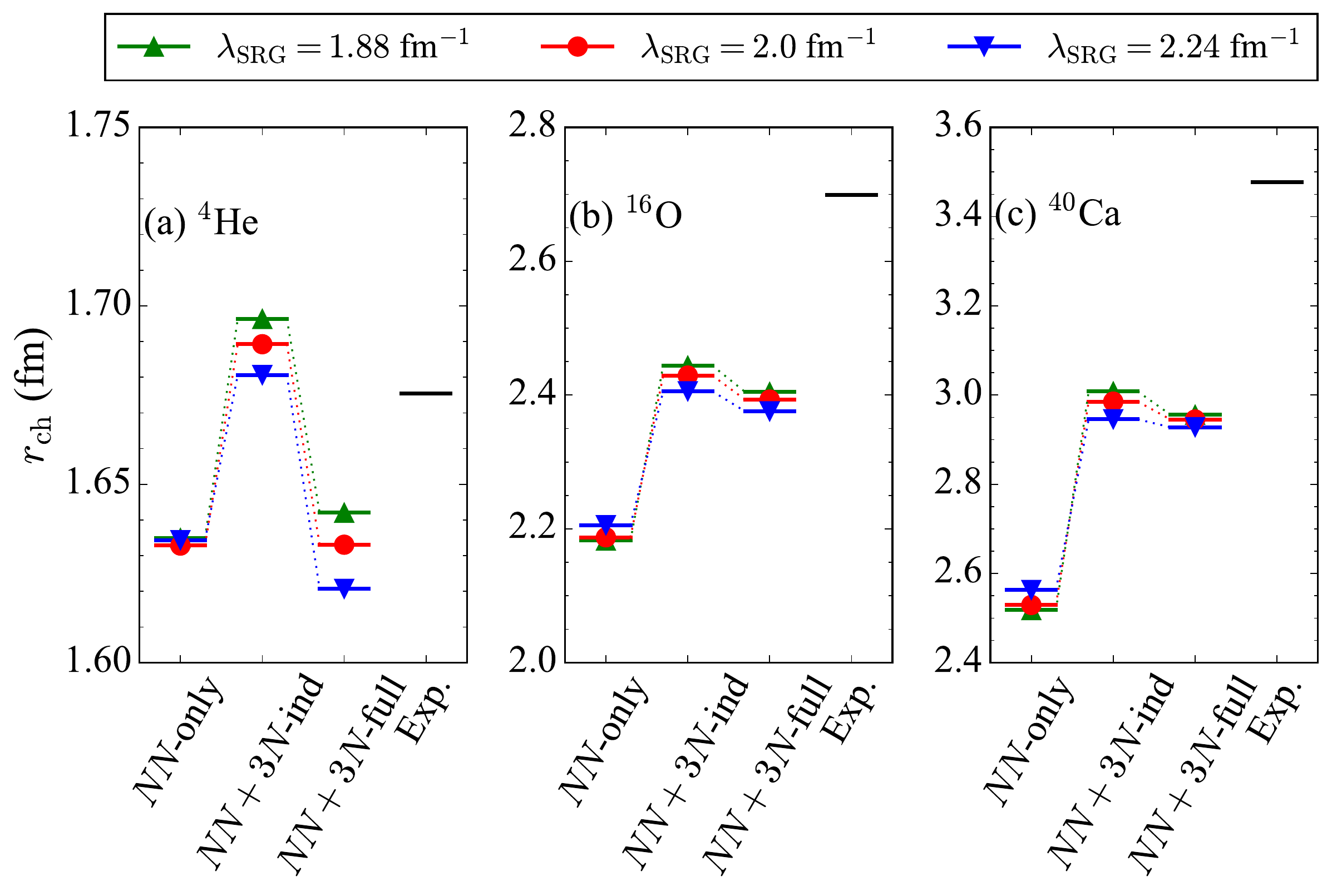}
  \caption{\label{fig:4} (color online) Charge radii for ${}^{4}$He, ${}^{16}$O, and
  ${}^{40}$Ca.
  All calculation results are obtained at $\hbar\omega = 25$ MeV and $e_{\max} = 14$.
  To evaluate the charge radii, the bare mean-squared radius operators are used.
  The experimental data are taken from Ref.~\cite{Angeli2013}.}
\end{figure}

\begin{figure}[t]
  \includegraphics[width=\columnwidth, clip]{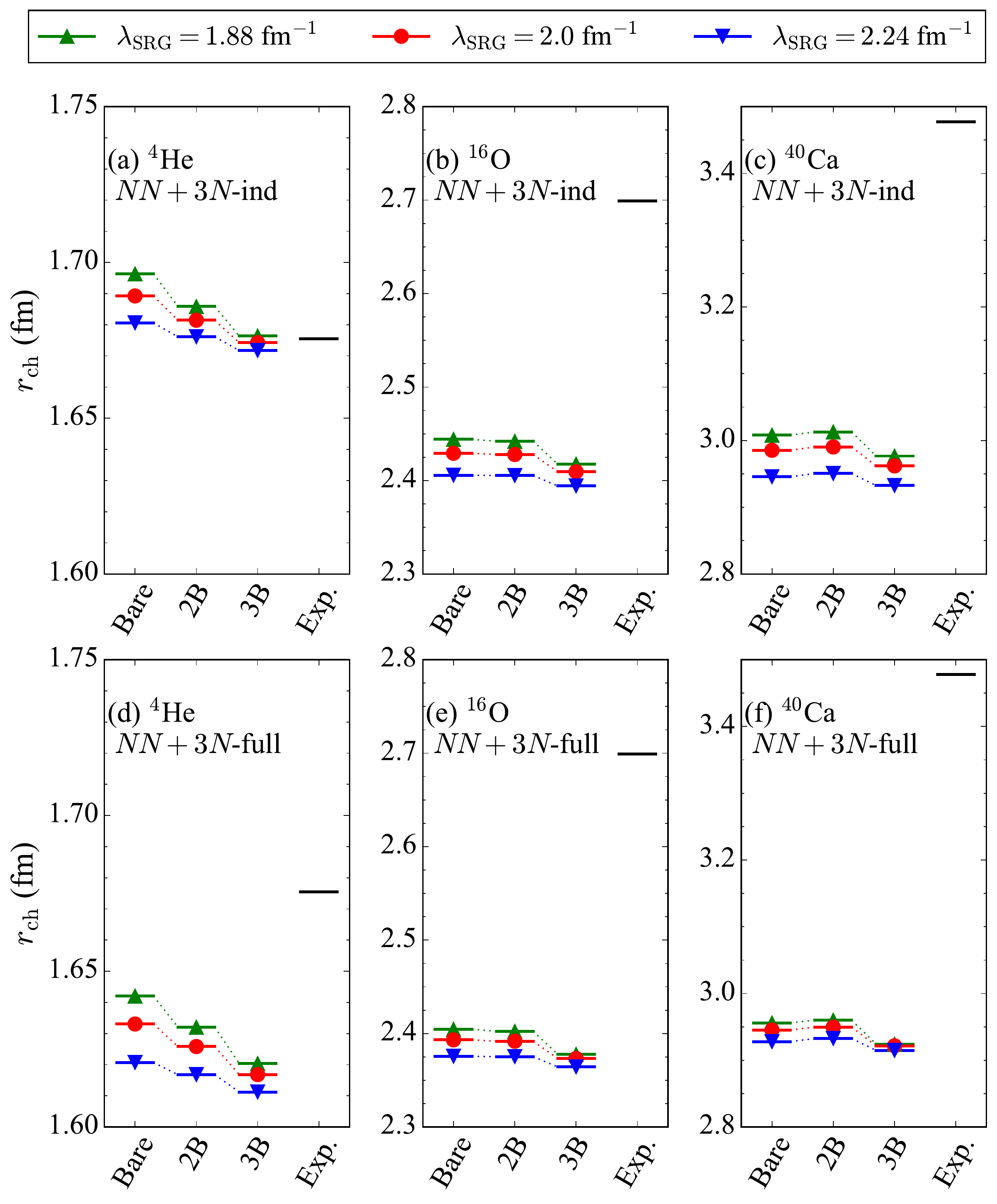}
  \caption{\label{fig:5} (color online) Charge radii evaluated with the expectation values
  of bare, two-body (2B) SRG evolved, and
  three-body (3B) SRG evolved mean-squared radius operator for
  ${}^{4}$He [(a) and (d)], ${}^{16}$O [(b) and (e)], and
  ${}^{40}$Ca [(c) and (f)]. All calculation results are
  obtained at $\hbar\omega = 25$ MeV and $e_{\max} = 14$.
  The calculation results in panels (a), (b), and (c) are calculated with $NN+3N$--ind interactions and
  the calculation results in panels (d), (e), and (f) are calculated with $NN+3N$--full interactions.}
\end{figure}
We also calculate radii for ${}^{4}$He, ${}^{16}$O, and ${}^{40}$Ca with the
 $NN$--only and $NN+3N$--ind
 interactions in the same manner as with the $NN+3N$--full interaction.
Then, we find converged results at $e_{\max} = 14$ and
 $\hbar\omega = 25$ MeV within $0.01$ fm.
The results are summarized in Fig.~\ref{fig:4} with the comparison to
 the experimental charge radii~\cite{Angeli2013}.
To compare with the experimental charge radii, our charge radii
 $r_{\rm ch}$ are evaluated as~\cite{Friar1975},
\begin{equation}
  \label{eq:rm2rch}
  r^{2}_{\rm ch} = r^{2}_{\rm g.s.} + r^{2}_{\rm p} +
  \frac{N}{Z}r^{2}_{\rm n} + \frac{3}{4m^{2}}.
\end{equation}
Here, we use $r_{\rm p} = 0.8751(61)$ fm~\cite{Patrignani2016},
 $r^{2}_{\rm n} = -0.1161(22)$ fm$^{2}$~\cite{Patrignani2016}, and
 $3/4m^{2} = 0.033$ fm$^{2}$, with the averaged nucleon mass $m = 938.919$ MeV$/c^{2}$.
Note that we assume the equivalence of point-proton and point-nucleon distributions in Eq.~(\ref{eq:rm2rch}).
This assumption would be reasonable because our targets are $N=Z$ stable nuclei.
In Fig.~\ref{fig:5}, the charge radii from $NN$--only interactions are obviously smaller than experimental data,
 especially for ${}^{16}$O and ${}^{40}$Ca and
 consistent with overbinding ground-state energies from those.
The SRG induced three-body operator contributes to spread the nuclear distribution out.
Then, the $\lambda_{\rm SRG}$-dependence is slightly enhanced.
With this particular Hamiltonian,
 the genuine $3N$ interaction shrinks nuclei and the calculated radii are
 clearly smaller than the experimental data.

As a possible reason for the calculated small radii, for example, the nuclear interaction can be
 considered.
In fact, the simultaneous reproduction of ground-state energies and radii were accomplished
 with the $\chi$EFT N$^{2}$LO $NN+3N$ interaction fitted by using some selected data of nuclei
 up to $A = 25$~\cite{Ekstrom2015}.
In addition, the saturation property of infinite nuclear matter was
 reproduced with the combinations of the softened N$^{3}$LO $NN$ and
 bare N$^{2}$LO $3N$ interactions whose low-energy constants are
 fitted to reproduce data of the few body systems~\cite{Hebeler2011}.
The simultaneous reproduction of ground-state energies and radii for finite nuclei with such
 interactions were compensated by the recent {\em ab initio} calculations~\cite{Simonis2017}.
As another possibility, we can consider amending the treatment of radius operator.
In earlier no-core shell model (NCSM) studies~\cite{Schuster2014, Schuster2015}, the effect of the SRG evolution
 to several operators were investigated for few-body systems.
However, such effects for medium-mass nuclei have not been clarified yet.
In this work, we investigate the effect of the SRG evolution to the radius operator.

\begin{table*}[]
  \caption{\label{tab3} Root-mean-squared radii $r_{\rm g.s.}$ for ${}^{4}$He, ${}^{16}$O, and ${}^{40}$Ca calculated
  with the bare, two-body evolved (2B), and three-body evolved (3B) radius operators.
  The results from both $NN$+$3N$--ind and $NN$+$3N$--full are displayed.
  All the calculation results are obtained at $e_{\max} = 14$ and $\hbar\omega = 25$ MeV.}
  \begin{ruledtabular}
  \begin{tabular}{clrrrrrr}
    & & \multicolumn{6}{c}{$r_{\rm g.s.}$ (fm)} \\ \cline{3-8}
    & & \multicolumn{3}{c}{$NN$+$3N$--ind} & \multicolumn{3}{c}{$NN$+$3N$--full} \\ \cline{3-8}
    Nuclide & $\lambda_{\rm SRG}$ (fm$^{-1}$) & Bare & 2B & 3B  & Bare & 2B & 3B \\ \hline
           & 1.88             &$1.48$  & $1.47$     &$ 1.46$   & $1.42$  & $1.40$     &$ 1.39$         \\
 ${}^{4}$He& 2.0              &$1.47$  & $1.46$     &$ 1.45$   & $1.41$  & $1.40$     &$ 1.39$         \\
           & 2.24             &$1.46$  & $1.45$     &$ 1.45$   & $1.39$  & $1.39$     &$ 1.38$         \\ \\
           & 1.88             &$2.32$  & $2.29$    &$2.26$   &   $2.28$  & $2.24$    &$2.22$         \\
 ${}^{16}$O& 2.0              &$2.30$  & $2.27$    &$2.25$   &   $2.27$  & $2.23$    &$2.21$         \\
           & 2.24             &$2.28$  & $2.25$    &$2.24$   &   $2.25$  & $2.22$    &$2.20$         \\ \\
           & 1.88             &$2.91$  & $2.89$    &$2.85$   &   $2.86$  & $2.83$    &$2.79$         \\
${}^{40}$Ca& 2.0              &$2.89$  & $2.86$    &$2.83$   &   $2.84$  & $2.82$    &$2.79$         \\
           & 2.24             &$2.84$  & $2.82$    &$2.80$   &   $2.83$  & $2.80$    &$2.78$         \\
 \end{tabular}
  \end{ruledtabular}
\end{table*}
We calculate the expectation value with the bare, two-body evolved, and three-body evolved
 mean-squared radius operators.
In the same as the bare radius operator cases, we check the convergence pattern and find the
 root-mean-squared radius results converged within $0.01$ fm.
Evaluated charge radii are illustrated in Fig.~\ref{fig:5}.
Corresponding to Fig.~\ref{fig:5}, final results for root-mean-squared radii
 from $NN+3N$--ind and $NN+3N$--full interactions are exhibited in
 Table~\ref{tab3}.
For all nuclei, as we calculate with higher-body evolved operator, the radii tend to be
 small and go opposite direction to the data.
This behavior is consistent with the earlier NCSM results~\cite{Schuster2014}.
Also, similar to the role of the SRG induced three-body interaction,
 consistently evolved operator moderately reduces the $\lambda_{\rm SRG}$-dependence of radii.
Therefore, it can be concluded that the consistent SRG evolution of the radius operator
 does not give the significant change compared to the experimental data.
This is consistent with the long-range nature of the radius operator~\cite{Schuster2014}.
There are some insights about the effect of consistent SRG evolution of radius operator~\cite{Miller2018}.
In this work, however, we do not observe the enhancement of radii discussed in Ref.~\cite{Miller2018}.
The quantitative reproduction of nuclear size is still an open question.

\begin{figure}[t]
  \includegraphics[width=\columnwidth, clip]{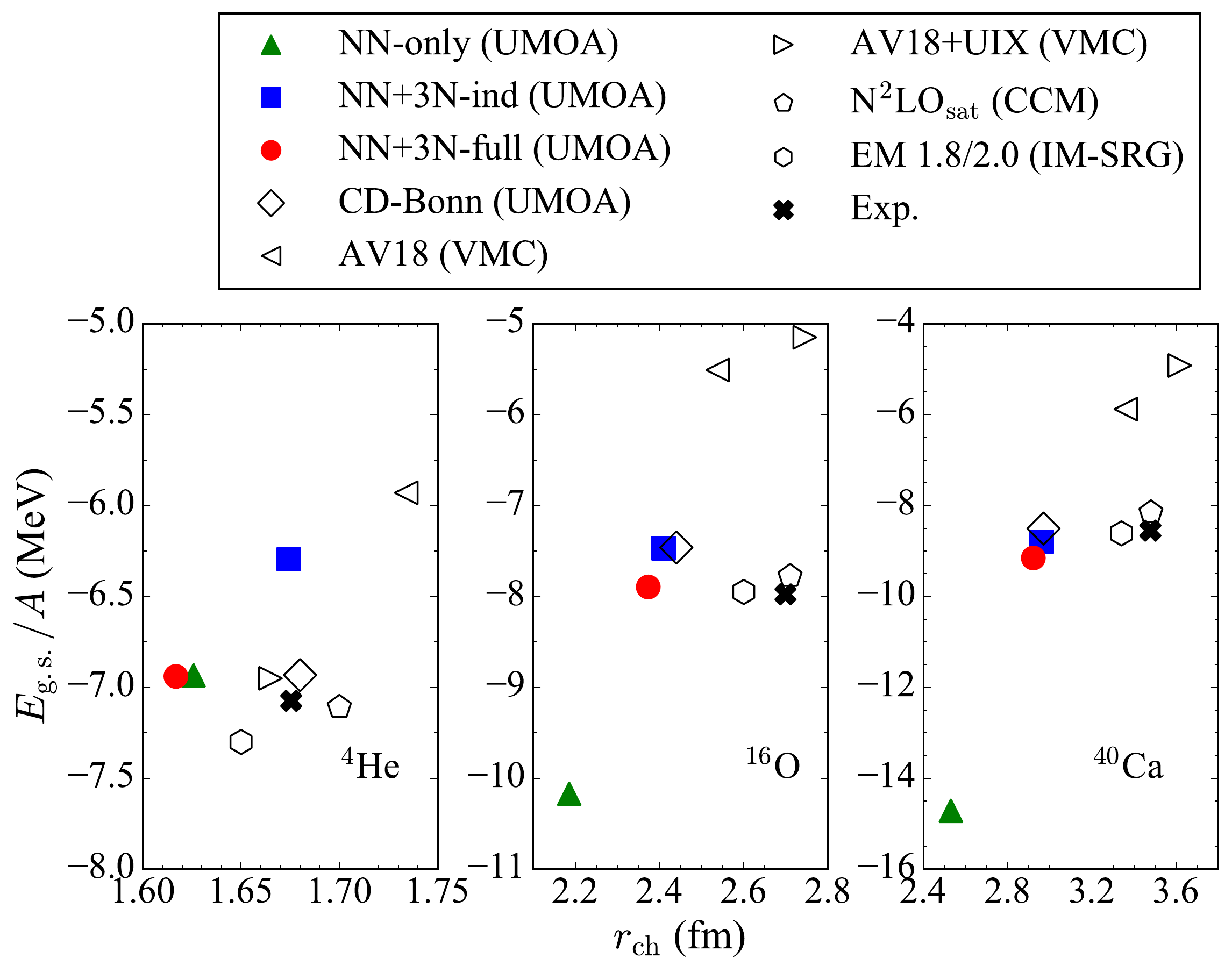}
  \caption{\label{fig:6} (color online)
  Ground-state energies per nucleon for ${}^{4}$He, $^{16}$O, and $^{40}$Ca as a function
   of $r_{\rm ch}$.
  The triangle, square, circle symbols are UMOA results from $NN$--only,
  $NN+3N$--ind, and $NN+3N$--full interactions evolved up to $2.0$ fm$^{-1}$, respectively.
  The diamonds are taken from earlier UMOA results with the CD-Bonn interaction~\cite{Miyagi2015a}.
  The left and right triangles are variational Monte Carlo calculations (VMC)
   with the AV18 and AV18+UIX interactions, respectively~\cite{Lonardoni2017}.
  The pentagons and hexagons are taken from the coupled-cluster method (CCM)~\cite{Ekstrom2015}
  and in-medium SRG (IM-SRG)~\cite{Simonis2017} results, respectively.
  Thick cross symbols indicate the experimental data~\cite{Wang2012,Angeli2013}.}
\end{figure}

Finally, let us see the saturation plot of the ground-state energies and
 radii from this work and from other calculations.
In Figure~\ref{fig:6}, the ground-state energies per nucleon
 and charge radii for $^{4}$He, $^{16}$O, and $^{40}$Ca with various nuclear interactions are plotted.
In the figure, the results obtained in the present work are shown by
 solid symbols.
The triangles, squares, and circles are the results with the $NN$--only,
 $NN+3N$--ind, and $NN+3N$--full interactions, respectively.
For visibility, only the results at $\lambda_{\rm SRG}=2.0$ fm$^{-1}$ are marked in the figure.
The open symbols are from other calculations~\cite{Miyagi2015a,Lonardoni2017,Ekstrom2015,Simonis2017}.
The experimental data are indicated by thick crosses~\cite{Wang2012,Angeli2013}.
The results with the $NN$ interactions, $NN$--only (triangle), $NN+3N$--ind (square),
 CD-Bonn (diamond), and AV18 (left triangle), fail to reproduce the experimental data.
The inclusion of 3N interactions, $NN$+3N full (circle) and
AV+UIX (right triangle), does not help the calculated radius come
close to the experimental data for $^{16}$O and $^{40}$Ca.
On the other hand, the other type of chiral $NN+3N$ interactions (pentagon and hexagon)
 show nice agreement with the data.
As seen in the figure, calculation results are scattered even if the $NN$ and $3N$ interactions are used, and
 further investigations are indispensable how both of the $NN$ and $3N$ interactions can be determined.

\section{Conclusion\label{sec:conc}}
In the present work, we have calculated the ground-state energies and radii for ${}^{4}$He,
 ${}^{16}$O, and ${}^{40}$Ca with the UMOA from $NN$ and $3N$ interactions based on
 the $\chi$EFT for the first time.
To obtain the computationally tractable Hamiltonian in the UMOA, we employed the SRG evolution
 and the NO2B approximation.

The resulting ground-state energies and radii agree with the recent {\em ab initio}
 calculation results within a few percent.
To discuss the accuracy of the UMOA calculation more precisely, we are going to
 extend the UMOA framework and directly treat the three-body cluster term
 beyond the NO2B approximation.
The results will be discussed in the future publication.

In addition to expectation values for the bare radius operator,
 in the present work, we have evaluated those
 for the two- and three-body SRG evolved radius operators.
By taking higher-body evolved operator into account, calculated radii slightly shrink, while
 the $\lambda_{\rm SRG}$-dependence of radii is reduced as we keep up to
 three-body terms.
Therefore, it is unlikely to reproduce the nuclear radii with the interactions employed
 in this work, even if we continue to include many-body terms induced by SRG evolution.
The simultaneous reproduction of the ground-state energies and radii strongly depend on empolyed nuclear interactions.
To specify the proper choice of nuclear interactions, further investigations are needed.

\begin{acknowledgments}
We thank R.~Roth for providing us the coupled-cluster calculation results.
This work was supported in part by JSPS KAKENHI Grant No. JP16J05707 and by the Program for Leading Graduate Schools, MEXT, Japan.
This work was also supported in part by MEXT as
 "Priority Issue on post-K computer" (Elucidation of the Fundamental Laws and Evolution of the Universe) and JICFuS
 (Projects No.~hp160211 and No.~hp170230) and
 CNS-RIKEN joint project for large-scale nuclear structure calculations, and
 the NSERC Grant No.~SAPIN-2016-00033.
TRIUMF receives federal funding via a
   contribution agreement with the National Research Council of Canada.
A part of numerical calculations were done on Oakforest-PACS, Supercomputing Division, Information Technology Center, the University of Tokyo.
\end{acknowledgments}

\end{document}